\documentclass[doublecol]{epl2}
\usepackage{graphicx}
\usepackage{amsmath}

\title{Fermi-edge problem in the presence of AC electric field}
\shorttitle{Title} 

\author{Yi Zhou\inst{1} \and Tai-Kai Ng\inst{1}}
\shortauthor{F. Author \etal}

\institute{
  \inst{1} Department of Physics, Hong Kong University of Science and
Technology, Clear Water Bay Road, Kowloon, Hong Kong
}

\abstract{
  We study in this paper a non-equilibrium Fermi-edge problem where the system under investigation is a single
  electron reservoir putting under an AC electric field. We show that the electron Green's function and other
  correlation functions in the problem can be solved and expressed exactly in terms of a well-defined integral.
  The qualitative behaviors of the solution is studied and compared with the situation where the impurity is
  coupled to more than one reservoir at different chemical potentials.}

\begin{document}

\maketitle

\begin{figure}[hpbt]
  \begin{center}
  \includegraphics[width=7.0cm]{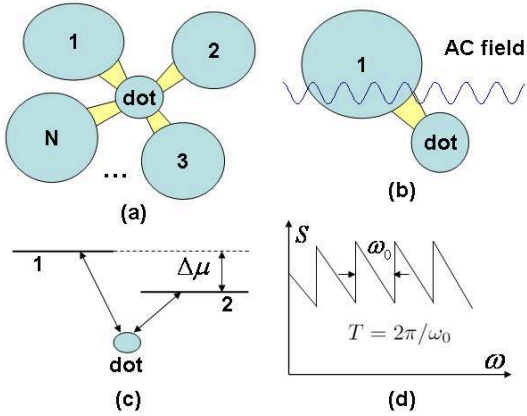}
  \end{center}
  \caption{(Color online) (a) $N$ reservoirs at different chemical potentials coupled to a quantum dot.
  (b) A single reservoir coupled to a quantum dot. The system is put under an AC electric field.
  (c) Dissipation caused by chemical potential differences rounds off the
  Fermi edge singularities. (d) Side-bands will appear for electrons in a single reservoir under an AC field.}
  \label{fig1}
  \end{figure}

     The Fermi edge problem in non-equilibrium situation is perhaps one of the most simple but non-trivial
  problems in non-equilibrium quantum many-body physics. It is the generalization of the Nozi$\acute{e}$res-De
  Dominicis X-ray edge problem\cite{ND} to non-equilibrium situations where the systems may be (i) connected to $N>1$
  reservoirs at different chemical potentials\cite{ng}, (ii) put under a time-dependent electromagnetic field, or
  (iii) put under other non-equilibrium configurations. The problem concerns the reaction of a non-equilibrium
  electron gas to the sudden turning on of a scattering potential. A simple realization of the problem is a small
  quantum dot coupled to electron reservoirs as shown in Fig.\ref{fig1}. The small quantum dot can be treated as a
  two-level system, and by changing the gate voltage applied to the quantum dot, the electron occupancy of the dot
  can be changed, resulting in a change of the local dot potential felt by electrons in the reservoirs.\cite{nf3,r1,r2}.
  Indeed, Fermi-edge singularity behavior was observed experimentally in a quantum dot system at energy range
  $\sim meV$\cite{r3}. The
  problem is in principle a one-particle problem where the many-body effects enter only through Fermi statistics.
  The problem has been studied carefully in those cases where the system is connected to reservoirs at different
  chemical potentials\cite{nf1,nf2,nf3,nf4}. It is now understood that in this case the problem is equivalent
  to solving an $N\times N$ matrix Riemann-Hilbert (RH) boundary value problem\cite{nf1,nf2} of which a general
  exact solution is unavailable and only the long-time asymptotic behavior of the solution can be
  extracted\cite{ng,nf1,nf2}.

   In this paper we consider a different situation where the system has only one electron reservoir putting under
  an AC electric field generated by, for example a microwave radiation. Going to the center of mass (CM) frame it
  can be shown that the electric field can be eliminated\cite{ng2} and the system becomes an ``equilibrium" system
  coupled to a time-dependent (periodic) impurity scattering potential. The corresponding Fermi-edge problem
  concerns the reaction of this ``equilibrium" electron gas to the sudden turning on of a time-dependent scattering
  potential. The problem provides a rare example of an non-trivial non-equilibrium quantum many-body problem which
  can be solved exactly and is testable. We find that the equation of motion for the electron Green's
  function is a Carleman type integral equation of which the solution can be expressed in terms of a well-defined
  integral. In this paper, the qualitative behaviors of the solution are analyzed in this paper and compared with the situation
  where the impurity potential is static but the system is connected to reservoirs at different chemical potentials.

  The system we consider is described by the time-dependent one-particle Hamiltonian $H=H_0(t)+V(t)$, where
  \begin{subequations}
  \begin{equation}
  \label{h1}
   H_0(t)=-\sum_{i}{\hbar^2\over2m}\nabla_i^2-e\sum_i\vec{r}_i\cdot\vec{E}(t)
  \end{equation}
  where the first term represents the kinetic energy of electrons, $\vec{E}(t)$ represents the AC electric field,
  $\vec{r}_i$ is the position of the $i^{th}$ electron and
  \begin{equation}
  \label{h2}
  V(t)=\theta(t_f-t)\theta(t-t_i)\sum_iU(\vec{r}_i)
  \end{equation}
  \end{subequations}
  describes a scattering potential $U(\vec{r})$ located at the origin turning on (off) at time $t=t_i (t_f)$.
  We have neglected spin as in usual practice.

   In the absence of $V(t)$ the electric field couples only to the CM coordinate $\vec{R}={1\over N}
  \sum_i\vec{r}_i$ of the whole electron fluid and can be eliminated by performing a coordinate transformation to
  CM frame of the system\cite{ng2}. After a gauge transformation, the Hamiltonian in CM
  frame becomes
  \begin{subequations}
  \label{hcm}
  \begin{equation}
  \label{hc1}
   H_{0(CM)}=\sum_{i}-{\hbar^2\over2m}\nabla_i^{'2}
  \end{equation}
  where $\vec{r}'_i=\vec{r}_i-\vec{R}(t)$, $\vec{R}(t)$ is the center of mass trajectory determined by the
  classical equation of motion $m\ddot{\vec{R}}=e\vec{E}(t)$. In the transformed coordinate, $V(t)$ becomes
  \begin{equation}
  \label{hc2}
  V_{CM}(t)=\theta(t_f-t)\theta(t-t_i)\sum_iU(\vec{r}'_i+\vec{R}(t)),
  \end{equation}
  \end{subequations}
  where the original static impurity becomes a moving impurity in the center of mass frame with corresponding
  time-dependent scattering potential $U_{CM}(\vec{r},t)=U(\vec{r}+\vec{R}(t))$.

    As in equilibrium situation we shall assume that the scattering events can be decomposed into separate
 scattering channels, and keep only one channel in our study. In this case we may replace the scattering potential
 by $U_{CM}(t)\sim U_0(t)\delta(\vec{r}'_i),$ where $U_0(t)$ is a periodic potential in time. With this
 approximation the only difference between the equilibrium Fermi edge problem and the present problem is the
 appearance of a time-dependent scattering potential $U_0(t)$. In realistic situations different scattering
 channels may be mixed by the travelling impurity. The qualitative behavior of the Fermi-edge singularity
 is not modified by this mixing\cite{nf2} as long as the scattering potentials in different channels have the
 same periodicity.

  Before presenting the solution we first revisit the case of a static impurity coupled to two reservoirs at
  different chemical potentials $\mu_1$ and $\mu_2$. In this case the chemical potential difference can be
  gauged away resulting in an effective time-dependent impurity scattering $V_{12}(t)\sim V_oe^{i\Delta\mu t}$
  ($\Delta\mu=\mu_1-\mu_2$) that couples the two reservoirs. The long-time ($t_f-t,t-t_i>>\hbar/\Delta\mu$)
  response of the electron gas to the suddenly switched on impurity potential is determined by two phase shifts
  associated with the two reservoirs. The phase shifts are complex when $\Delta\mu\neq0$ and reflect the
  existence of dissipation in an out-of-equilibrium system. Associated with the two phase shifts are two
  Fermi-edge singularities which are rounded off by finite life-time effects associated with the imaginary part of
  the complex phase shifts\cite{ng,nf1,nf2,nf3}. For electrons in a single reservoir under a time-dependent
  electric field with period $T=2\pi/\omega_0$, side-bands will appear in the electron spectral function at
  frequencies $\omega\sim\mu+n\omega_0$, where $n=$ integer. An impurity will cause scattering between different
  side-bands. Will Fermi edge singularities appear on all these side-bands and will the singularities be rounded
  off by finite-life time effect as in the many-reservoir situation? We note that both AC and DC transports are
  usually dissipative in condensed matter systems. Will there be other new effects?
  These are questions to be asked in the present problem.

  At zero temperature, the main challenge of the Fermi-edge problem is to evaluate the one-particle
  time-ordered Green's function $G(t,t')$ satisfying the Dyson equation
  \begin{subequations}
  \label{dyson}
  \begin{equation}
  G(t,t')=g(t-t')+i\int^{t_f}_{t_i}g(t-t")U_0(t")G(t",t')dt",
  \end{equation}
  where
  \begin{equation}
  \label{green}
  g(t-t')\sim-i\nu_0\left(P{1\over{t-t'}}+\pi\tan\theta_0\delta(t-t')\right)
  \end{equation}
  \end{subequations}
  is the unperturbed Green's function determined by $H_{0}$. The Green's functions are defined in the CM
  coordinate in Eq.\ (\ref{dyson}). A similar equation can also be derived at finite temperature\cite{nf4} with
  $g$ replaced by its finite temperature extension. Since the Hamiltonian is quadratic, other correlation
  functions can be computed once $G$ is determined. An important correlation function characterizing the Fermi-edge
  singularity is the propagator
  $$
  B(t_f,t_i)=e^{C(t_f,t_i)}=\left\langle\psi_0\right|S(t_f,t_i)\left|\psi_0\right\rangle,
  $$
  where $S(t_f,t_i)=T\exp(-{i\over\hbar}\int^{t_f}_{t_i}U_0(t')dt')$ is the time evolution operator under the
  scattering potential $U_0(t)$ and $\left|\psi_0\right\rangle$ is the ground state wavefunction of the system
  (in the CM frame) before $U_0(t)$ is switched on. $C(t_f,t_i)$ measures the overlap between
  $\left|\psi_0\right\rangle$ and the final state after $U_0(t)$ is switched on (orthogonality
  catastrophe\cite{and}) and is related to $G$ by
  \begin{equation}
  \label{cG}
  \lambda{\partial C(t_f,t_i)\over\partial\lambda}=i\int^{t_f}_{t_i}(\lambda U_0(t))G^{\lambda}(t,t)dt
  \end{equation}
  where $G^{\lambda}$ is the Green's function determined from Eq.\ (\ref{dyson}) with $U_0(t)\to\lambda
  U_0(t)$\cite{ND}. The electron Green's function measuring the Fermi-edge singularity is\cite{ND}
  $$
  K(t_f,t_i)=G(t\to t_f,t'\to t_i)e^{C(t_f,t_i)}.
  $$

    To solve Eq.\ (\ref{dyson}) we introduce $\tilde{G}(t,t')=\nu_0U_0(t)G(t,t')$. After some simple
    manipulation we obtain
  \label{carleman}
  \begin{equation}
  \label{car1}
  \alpha(t)\tilde{G}(t,t')=g(t-t')+P\int^{t_f}_{t_i}{\tilde{G}(t",t')\over t-t"}dt",
  \end{equation}
  where $\alpha(t)={1-\pi\nu_0\tan\theta_0U_0(t)\over\nu_0U_0(t)}$.
  The equation is a Carleman type equation with $t'$ as a dummy variable. The solution to this equation
  can be constructed by analyzing the analytic behavior of the equation on the complex
  plane\cite{math} (essentially a RH problem) and the relevant solution to our problem is
  \begin{subequations}
  \label{solution}
  \begin{eqnarray}
  \tilde{G}(t,t')&=&{\sin\delta(t)\over\pi}\times{\Bigl[}\cos\delta(t)g(t-t')\label{sol1}\\
  &&+{e^{-\varphi(t)}\over\pi}P\int_{t_i}^{t_f}{e^{\varphi(t")}g(t"-t')\sin\delta(t")\over{t-t"}}dt"{\Bigr]}\nonumber
  \end{eqnarray}
  where $\delta(t)=\tan^{-1}(\pi/\alpha(t))$ and
  \begin{equation}
  \label{sol2}
  \varphi(t)=Re\Phi(t+i\epsilon)=-{1\over\pi}P\int^{t_f}_{t_i}{\delta(t')\over t-t'}dt'.
  \end{equation}
  \end{subequations}
  Notice that the complex function $\Phi(z)=-{1\over\pi}\int^{t_f}_{t_i}{\delta(t')\over z-t'}dt'$ is an
  analytical function on the complex plane except the segment $[t_i,t_f]$. Using Eq.\ (\ref{green})
  for $g$ and the fact that
 $$
 -{1\over\pi}P\int^{t_f}_{t_i}{1\over t-t'}e^{\varphi(t')}\sin\delta(t')dt'=e^{\varphi(t)}\cos\delta(t)
 $$
 at time $t_i<t<t_f$ because the integral is just the Kramers-Kronig relation relating the real and imaginary parts
 of the analytic function $\textbf{F}(t+i\epsilon)=e^{\Phi(t+i\epsilon)}$, we obtain after some algebra,
  \begin{eqnarray}
  \label{sol3}
  G(t,t')&=&(-i\nu_0)A_0(t)\times\bigl[e^{-\varphi(t)}e^{\varphi(t')}A_0(t')P{1\over{t-t'}}\nonumber\\
  &&+\left(\cos\delta(t)\tan\theta_0-\sin\delta(t)\right)\pi\delta(t-t')\bigr]
  \end{eqnarray}
  for $t_i<t,t'<t_f$, where $A_0(t)=(1+\tan\delta(t)\tan\theta_0)\cos\delta(t)$.

    First we examine eq.\ (\ref{sol3}) in the equilibrium situation
    $U_0(t)=U_0$. In this case
  $\delta(t)\to\delta$ becomes a constant and $\varphi(t)\to{\delta\over\pi}ln\left({t_f-t\over t-t_i}\right)$.
  It is easy to see that the ND result for the electron Green's function\cite{ND} is recovered. The non-trivial
  result here is that the phase shift $\delta(t)$ becomes time-dependent because of the
  time-dependent potential $U_0(t)$. Notice that contrary to the case of a static impurity coupled to reservoirs
  at different chemical potentials where the relevant phase shifts are complex, the phase shift $\delta(t)$ here
  is always {\em real}.

  It is straightforward to show that
  $$
  \tilde{G}(t,t'\to t)\to(-i\nu_0){\sin\delta(t)\over\pi}A_0(t)
  \left(i\epsilon_c-{\partial\varphi(t)\over\partial t}\right),
  $$
  where $i\epsilon_c\sim\lim_{t\to t'}{1\over t-t'}$ is a high-energy
  cutoff below which the approximate expression\ (\ref{green}) for $g_0$ is valid\cite{ND}.
  Putting this back into Eq.\ (\ref{cG}) and using the identity $d\lambda/\lambda=d\delta/(A_0\sin\delta)$ we
  obtain after some simple algebra
  \begin{equation}
  C(t_f,t_i)=
  \int^{t_f}_{t_i}dt\int^{\delta(t)}_0{d\delta\over\pi}\left(i\epsilon_c-
  {\partial\varphi(t)\over\partial t}\right).  \label{cf}
  \end{equation}

    In the equilibrium case $\delta(t)=\delta$, the first term $\sim\epsilon_c\int^{t_f}_{t_i}dt\delta(t)$
  in Eq.\ (\ref{cf}) represents a self-energy correction to ground state energy from impurity scattering
  (Fumi's Theorem\cite{fumi}). The second term introduces logarithmic corrections to $C(t_f,t_i)$ coming from
  suddenly switching on(off) the scattering potentials at $t_i(t_f$) (orthogonality catastrophe)\cite{ND,and}.
  Eq.\ (\ref{cf}) represents a generalization of this result to time-dependent scattering potentials. We shall
  consider the situation where $\alpha(t)$ is periodic in time in the following. In this case, $\delta(t)$ is also
  periodic in time and we can write $\delta(t)=\delta_0+\sum_{n\neq0}\delta_ne^{in\omega_0t}$,
  where $\delta_n^*=\delta_{-n}$ ($\delta(t)$ is real).

  $\varphi(t)$ cannot be evaluated exactly in this case. However it is easy to extract the short- and long-time
  behaviors of $\varphi(t)$. In the short-time limit $\epsilon_c^{-1}<<t-t_i,t_f-t<<T$ the leading contribution is
  \begin{equation}
  \label{short}
  \varphi(t)\sim{\delta(t)\over\pi} ln\left({t_f-t\over t-t_i}\right)
  \end{equation}
  whereas in the opposite limit $t-t_i,t_f-t>>T$ we obtain
  \begin{subequations}
  \label{phi}
  \begin{equation}
  \label{phi1}
  \varphi(t)={\delta_0\over\pi}ln({t_f-t\over t-t_i})+\varphi_{\bar{n}}(t),
  \end{equation}
  where $\varphi_{\bar{n}}(t)=\varphi^{\infty}(t)+\varphi_c(t)$ is the contribution from $\delta_{n\neq0}$ terms
  in the Fourier representation of $\delta(t)$. $\varphi^{\infty}(t)=i\sum_{n\neq0}\delta_n sig(n)e^{in\omega_0t}$
  is the contribution in the limit $-t_i=t_f\to\infty$ and
  \begin{eqnarray}
  \label{phi3}
  \varphi_c(t)&=&-i\sum_{n\neq0}{\delta_n\over\pi}{\Bigl[}e^{in\omega_0t_f}Ei(n\omega_0(t_f-t))\nonumber\\
  &&-e^{in\omega_0t_i}Ei(n\omega_0(t_i-t){\Bigr]}
  \end{eqnarray}
  \end{subequations}
  is the correction from finite time cutoff $t_i (t_f)$. $Ei(x)=P\int_{-\infty}^{x}{e^{-y}\over y}dy$ is the
  exponential integral function. It is easy to see that $\varphi^{\infty}(t)$ is a real and periodic function in
  time. Using the asymptotic results $Ei(z)\sim lnz$ for $|z|<<1$ and $Ei(z)\sim e^z/z$ for $|z|>>1$ we find that
  $\varphi_c(t_{f(i)})\sim b(t_{f(i)})+{i\over(t_f-t_i)}\int^{t_{i(f)}}dt'\delta(t')$ in the limit $t_f-t_i>>T$
  where $b(t)$ is a periodic function in time and the second term is small compared with
  the $\delta_0$ term in Eq.\ (\ref{phi1}).

    With these results we can now evaluate the Green's function $G(t\to t_f,t'\to t_i)$ in
  the long and short time limits $t_f-t_i>>T$ and $\epsilon_c^{-1}<<t_f-t_i<<T$. In the long time limit we obtain
  \begin{equation}
  \label{G1}
  G(t\to t_f,t'\to t_i)\to{-i\nu_0\over
  t_f-t_i}A_-(t_f)\left(i\epsilon_c(t_f-t_i)\right)^{2\delta_0\over\pi}A_+(t_i)
  \end{equation}
  where $A_{\pm}(t)=A_0(t)e^{\mp\varphi_{\bar{n}}(t)}$ are periodic functions in time. The corresponding
  Green's function in the short-time limit can be obtained by replacing $2\delta_0$ by
  $\delta(t_i)+\delta(t_f)\sim2\delta((t_i+t_f)/2)$ and $A_{\pm}(t)$ by $A_0(t)$ in Eq.\ (\ref{G1}).

  It is also straightforward to obtain in the long time limit $t_f-t_i>>T$
  \begin{equation}
  \label{cf1}
  C(t_f,t_i)=i\int^{t_f}_{t_i}dt{\delta(t)\over\pi}\epsilon_c-{\delta_0^2\over\pi^2}ln(i\epsilon_c(t_f-t_i))
  +\bar{\varphi}(t_f)-\bar{\varphi}(t_i)
  \end{equation}
  where $\bar{\varphi}(t)$
   is a real and periodic function of time $t$ to leading order in $T/(t_f-t_i)$ coming from
   the $\varphi_{\bar{n}}(t)$ term. The short-time behavior of
   $C(t_f,t_i)$ can be obtained by replacing $\delta_0^2$ by $(\delta(t_f)^2+\delta(t_i)^2)/2\sim
   \delta((t_i+t_f)/2)^2$ and dropping the $\bar{\varphi}(t)$ terms in Eq.\ (\ref{cf1}).

   Combining these results we find that the electron Green's function $K=Ge^C$
   can be written as
   \begin{equation}
   \label{ff}
   K(t_f,t_i)\sim K^0(t_f,t_i)\left(i\epsilon_c(t_f-t_i)\right)^{\gamma((t_f+t_i)/2)}
   \end{equation}
   in both the long and short time limits where  $\gamma(t)\sim2\delta(t)/\pi-\delta(t)^2/\pi^2$
   in the short-time limit $\epsilon_c^{-1}<<t_f-t_i<<T$ and $\gamma(t)\rightarrow\gamma_0=
   2\delta_0/\pi-\delta_0^2/\pi^2$ in the long-time limit $t_f-t_i>>T$.
   $K^0(t,t')=\bar{A}_-(t)P{1\over t-t'}\bar{A}_+(t')$ where $\bar{A}_{\pm}(t)$ are
   periodic (but different) functions in time in the long and short time limits.

    With Eq.\ (\ref{ff}) we can evaluate the Fourier transformed Green's function
    $K(\omega,\omega')=\int dt\int dt'e^{i(\omega t-\omega't')}K(t,t')\theta(t-t')$ in
    the frequency regimes $\epsilon_c>>|\omega+\omega'|>>2\omega_0$ and $|\omega+\omega'|<<2\omega_0$.
    In the low frequency limit $|\omega+\omega'|<<2\omega_0$, we obtain
    \begin{equation}
     \label{klong}
     K(\omega,\omega')\sim\sum_n\delta(\omega-\omega'+n\omega_0)\bar{\alpha}_n
     \left({\epsilon_c\over\omega+n\omega_0/2}\right)^{\gamma_0},
     \end{equation}
     where $\bar{\alpha}_n$ are coefficients which depends on the strength and polarization of the AC electric field.

    In the high-frequency regime $2\epsilon_c>>|\omega+\omega'|>>2\omega_0$, $K(\omega,\omega') \sim
    \int_{-\infty}^{\infty}dt_m e^{i(\omega-\omega')t_m}A_0(t_m)^2 \left({2\epsilon_c\over\omega+\omega'}
    \right)^{\gamma(t_m)}$. Evaluating the integral we obtain
    \begin{equation}
    \label{kshort}
    K(\omega,\omega') \sim  \sum_n\delta(\omega-\omega'+n\omega_0)\alpha_n(\omega+\omega')
    \left({\epsilon_c\over\omega+n\omega_0/2}\right)^{\gamma_s},
    \end{equation}
     where $\gamma_s\sim\langle\gamma(t_m)\rangle$ is the critical exponent in the corresponding equilibrium Fermi-edge problem.
     $\alpha_n(\omega+\omega')$ are constants up to logarithmic correction factors. The Fermi-edge singularity in the
     high-frequency regime $|\omega+\omega'|>>2\omega_0$ is insensitive to the AC electric field. Notice
     that $\alpha_n\neq\bar{\alpha}_n$ and $\gamma_0\neq\gamma_s$ in general.

    Equations (\ref{klong}) and (\ref{kshort}) predict that the Fermi-edge spectrum $Re K(\omega,\omega')=
    \sum_n\delta(\omega-\omega'+n\omega_0)S_n(\omega)$ develops side bands as a result of the time-periodic
    potential. The spectral functions $S_n(\omega)$ develop power-law singularities at $\omega\rightarrow-
    n\omega_0/2$ characterized by an $n$-independent Fermi-edge exponent $\gamma_0$ for
    $\omega+n\omega_0/2<<\omega_0$. The Fermi edge singularities are {\em not} round-off by finite-life time
    effects as in the case of systems with reservoirs at different chemical potentials. This is non-trivial
    since dissipation usually exists in both DC and AC transports in many-reservoir problems\cite{ng,tdimpurity}.
    In fact, it was shown in Ref.\cite{tdimpurity} that dissipation exists in a system of quantum dot coupled to external
    reservoirs when the system is driven by a frequency $f<<W$, where $W$ is the bandwidth of the quantum dot
    (quantum pump problem). The finite quantum dot bandwidth provides a natural mechanism for dissipation. Our system
    which consists of only one single impurity corresponds roughly to the opposite limit $f>>W\rightarrow0$ of
    Ref.\cite{tdimpurity} where a natural mechanism for dissipation does not exist. The Fermi-edge singularity at the
    high-frequency regime $\epsilon_c>>\omega>>\omega_0$ is found to be characterized by an envelope
    function covering all side-bands with an exponent $\gamma_s$ which is insensitive to AC field and
    is different from the low frequency exponent $\gamma_0$ (see figure 2). These predicted features
    can be tested in mesoscopic systems under AC electric fields.

   \begin{figure}[hpbt]
   \begin{center}
   \includegraphics[width=7.5cm]{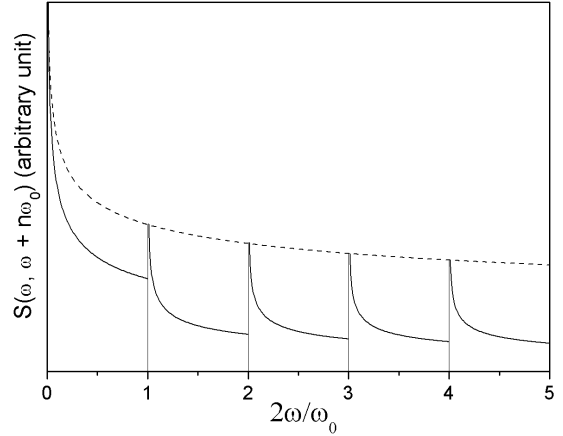}
   \end{center}
   \caption{Spectral function in presence of AC field.}
   \label{fig2}
   \end{figure}

     Summarizing, we analyze in this paper a non-equilibrium Fermi-edge problem where the system under
   consideration is driven out-of-equilibrium by a time-dependent (AC) electric field. We show that the system can be
   mapped into an equilibrium system with a travelling impurity, or an impurity with a time-dependent scattering
   potential. We find that the problem can be solved exactly and the corresponding Fermi-edge singularity behavior is
   very different from the case of a static impurity coupled to many reservoirs at different chemical potentials. Our
   solution provides a rare example of
   non-equilibrium quantum many-body problem which can be solved exactly.

  \acknowledgements
  We acknowledge support from HKUGC through grant 602803.


\begin{thebibliography}{0}

 \bibitem{ND}
   \Name{Nozi$\grave{e}$res P. \and De Dominicis C.T.}
   \REVIEW{Phys. Rev.}{178}{1969}{1097}.

 \bibitem{ng}
   \Name{Ng T.K.}
   \REVIEW{Phys. Rev. B}{54}{1996}{5814}.

 \bibitem{nf3}
   \Name{Abanin D.A. \and Levitov L.S.}
   \REVIEW{Phys. Rev. Lett.}{93}{2004}{126802}.

 \bibitem{r1}
   \Name{ Matveev K.A. \and Larkin A.I.}
   \REVIEW{Phys. Rev. B}{46}{1992}{15337}.

 \bibitem{r2}
   \Name{Geim A.K. {\em et. al.}}
   \REVIEW{Phy. Rev. Lett.}{72}{1994}{2061}).

 \bibitem{r3}
   \Name{Maire N. {\em et. al.}}
   \REVIEW{Phys. Rev. B}{75}{2007}{233304}.

 \bibitem{nf1}
   \Name{Muzykantskii B. \and d'Ambrumenil N. \and Braunecker B.}
   \REVIEW{Phy. Rev. Lett.}{91}{2003}{266602}.

 \bibitem{nf2}
   \Name{d'Ambrumenil N. \and Muzykantskii B.}
   \REVIEW{Phys. Rev. B}{71}{2005}{045326}.

 \bibitem{nf4}
   \Name{Braunecker B.}
   \REVIEW{Phys. Rev. B}{73}{2006}{075122}.

 \bibitem{ng2}
   \Name{Ng T.K.\and Dai L.X.}
   \REVIEW{Phys. Rev. B}{72}{2005}{235333}.

 \bibitem{and}
   \Name{Anderson P.W.}
   \REVIEW{Phys. Rev. Lett.}{18}{1967}{1049}.

 \bibitem{math}
   \Name{Estrada R. \and Kanwal R.P.}
   \Book{Singular integral equations}
   \Publ{Birkh\"{a}user, Boston}
   \Year{2000}.

 \bibitem{fumi}
   \Name{Fumi F.G.}
   \REVIEW{Philos. Mag.}{46}{1955}{1007}.

 \bibitem{tdimpurity}
   \Name{Avron J. E. \and Elgart A. \and Graf G. M. \and Sadun L.}
   \REVIEW{Phys. Rev. Lett.}{87}{236601}{2001};
   \Name{Arrachea L. \and Moskalets L. \and Martin-Moreno L.}
   \REVIEW{Phys. Rev. B}{75}{245420}{2007}.

\end{thebibliography}
\end{document}